\documentclass[prl,superscriptaddress,showpacs,twocolumn,amssymb,amsmath,amsfonts,aps]{revtex4}

\usepackage{graphicx}% Include figure files
\usepackage{dcolumn}% Align table columns on decimal point
\usepackage{bm}% bold math

\begin{document}

\preprint{CMU/JLab/5-2003}

%%%%%%%%%%%%%%%%% % List of authors
% Generated from the clas database. 
% Concerns about content related to future author lists email clasmbr@jlab.org 
% Concerns about content related to this publication must be sent to first authors
% %%%%%%%%%%%%%%%%%%%%%%%%%%%%%%%%%%% % commands
%   define a shorthand notation XXX for affiliation names and addresses  
%      \newcommand*{\XXX}{Name, Address}
%   use the new commands
%        \affiliation{\XXX}     standard 
%        \altaffiliation{\XXX}  current address if different from above 
% to add the University and address information to each author
%
%    start boundary 1=    photohyperon_prl.aux      Sun Jan 31 00:00:52 EST 1999
%    end boundary 1  =          Sat May 01 00:00:52 EDT 1999
%
%  Photon only people have been excluded %   Electron only data assumed %%
%
% FIRST time through establishes the order
%%%%% 
\newcommand*{\CMU }{ Carnegie Mellon University, Pittsburgh, Pennsylvania 15213} \affiliation{\CMU } 
\newcommand*{\ASU }{ Arizona State University, Tempe, Arizona 85287-1504} \affiliation{\ASU } 
\newcommand*{\SACLAY }{ CEA-Saclay, Service de Physique Nucl\'eaire, F91191 Gif-sur-Yvette, Cedex, France} \affiliation{\SACLAY } 
\newcommand*{\UCLA }{ University of California at Los Angeles, Los Angeles, California  90095-1547} \affiliation{\UCLA } 
\newcommand*{\CUA }{ Catholic University of America, Washington, D.C. 20064} \affiliation{\CUA } 
\newcommand*{\CNU }{ Christopher Newport University, Newport News, Virginia 23606} \affiliation{\CNU } 
\newcommand*{\UCONN }{ University of Connecticut, Storrs, Connecticut 06269} \affiliation{\UCONN } 
\newcommand*{\DUKE }{ Duke University, Durham, North Carolina 27708-0305} \affiliation{\DUKE } 
\newcommand*{\GBEDINBURGH }{ Edinburgh University, Edinburgh EH9 3JZ, United Kingdom} \affiliation{\GBEDINBURGH } 
\newcommand*{\FIU }{ Florida International University, Miami, Florida 33199} \affiliation{\FIU } 
\newcommand*{\FSU }{ Florida State University, Tallahassee, Florida 32306} \affiliation{\FSU } 
\newcommand*{\GWU }{ The George Washington University, Washington, DC 20052} \affiliation{\GWU } 
\newcommand*{\GBGLASGOW }{ University of Glasgow, Glasgow G12 8QQ, United Kingdom} \affiliation{\GBGLASGOW } 
\newcommand*{\INFNFR }{ INFN, Laboratori Nazionali di Frascati, Frascati, Italy} \affiliation{\INFNFR } 
\newcommand*{\INFNGE }{ INFN, Sezione di Genova, 16146 Genova, Italy} \affiliation{\INFNGE } 
\newcommand*{\ORSAY }{ Institut de Physique Nucleaire ORSAY, Orsay, France} \affiliation{\ORSAY } 
\newcommand*{\ITEP }{ Institute of Theoretical and Experimental Physics, Moscow, 117259, Russia} \affiliation{\ITEP } 
\newcommand*{\JMU }{ James Madison University, Harrisonburg, Virginia 22807} \affiliation{\JMU } 
\newcommand*{\KYUNGPOOK }{ Kyungpook National University, Daegu 702-701, South Korea} \affiliation{\KYUNGPOOK } 
\newcommand*{\MIT }{ Massachusetts Institute of Technology, Cambridge, Massachusetts  02139-4307} \affiliation{\MIT } 
\newcommand*{\UMASS }{ University of Massachusetts, Amherst, Massachusetts  01003} \affiliation{\UMASS } 
\newcommand*{\MOSCOW }{ Moscow State University, 119899 Moscow, Russia} \affiliation{\MOSCOW } 
\newcommand*{\UNH }{ University of New Hampshire, Durham, New Hampshire 03824-3568} \affiliation{\UNH } 
\newcommand*{\NSU }{ Norfolk State University, Norfolk, Virginia 23504} \affiliation{\NSU } 
\newcommand*{\OHIOU }{ Ohio University, Athens, Ohio  45701} \affiliation{\OHIOU } 
\newcommand*{\ODU }{ Old Dominion University, Norfolk, Virginia 23529} \affiliation{\ODU } 
\newcommand*{\PENN }{ Penn State University, University Park, Pennsylvania 16802} \affiliation{\PENN } 
\newcommand*{\PITT }{ University of Pittsburgh, Pittsburgh, Pennsylvania 15260} \affiliation{\PITT } 
\newcommand*{\ROMA }{ Universita' di ROMA III, 00146 Roma, Italy} \affiliation{\ROMA } 
\newcommand*{\RPI }{ Rensselaer Polytechnic Institute, Troy, New York 12180-3590} \affiliation{\RPI } 
\newcommand*{\RICE }{ Rice University, Houston, Texas 77005-1892} \affiliation{\RICE } 
\newcommand*{\URICH }{ University of Richmond, Richmond, Virginia 23173} \affiliation{\URICH } 
\newcommand*{\SCAROLINA }{ University of South Carolina, Columbia, South Carolina 29208} \affiliation{\SCAROLINA } 
\newcommand*{\UTEP }{ University of Texas at El Paso, El Paso, Texas 79968} \affiliation{\UTEP } 
\newcommand*{\JLAB }{ Thomas Jefferson National Accelerator Facility, Newport News, Virginia 23606} \affiliation{\JLAB } 
\newcommand*{\UNIONC }{ Union College, Schenectady, NY 12308} \affiliation{\UNIONC } 
\newcommand*{\VT }{ Virginia Polytechnic Institute and State University, Blacksburg, Virginia   24061-0435} \affiliation{\VT } 
\newcommand*{\VIRGINIA }{ University of Virginia, Charlottesville, Virginia 22901} \affiliation{\VIRGINIA } 
\newcommand*{\WM }{ College of Willliam and Mary, Williamsburg, Virginia 23187-8795} \affiliation{\WM } 
\newcommand*{\YEREVAN }{ Yerevan Physics Institute, 375036 Yerevan, Armenia} \affiliation{\YEREVAN } 
\newcommand*{\NOWNCATU }{ North Carolina Agricultural and Technical State University, Greensboro, NC 27411}
\newcommand*{\NOWGBGLASGOW }{ University of Glasgow, Glasgow G12 8QQ, United Kingdom}
\newcommand*{\NOWJLAB }{ Thomas Jefferson National Accelerator Facility, Newport News, Virginia 23606}
\newcommand*{\NOWSCAROLINA }{ University of South Carolina, Columbia, South Carolina 29208}
\newcommand*{\NOWFIU }{ Florida International University, Miami, Florida 33199}
\newcommand*{\NOWINFNFR }{ INFN, Laboratori Nazionali di Frascati, Frascati, Italy}
\newcommand*{\NOWCMU }{ Carnegie Mellon University, Pittsburgh, Pennsylvania 15213}
\newcommand*{\NOWINDSTRA }{ Systems Planning and Analysis, Alexandria, Virginia 22311}
\newcommand*{\NOWASU }{ Arizona State University, Tempe, Arizona 85287-1504}
\newcommand*{\NOWCISCO }{ Cisco, Washington, DC 20052}
\newcommand*{\NOWUK }{ Kentucky, LEXINGTON, KENTUCKY 40506}
\newcommand*{\NOWSACLAY }{ CEA-Saclay, Service de Physique Nucl\'eaire, F91191 Gif-sur-Yvette, Cedex, France}
\newcommand*{\NOWRPI }{ Rensselaer Polytechnic Institute, Troy, New York 12180-3590}
\newcommand*{\NOWUNCW }{ North Carolina}
\newcommand*{\NOWHAMPTON }{ Hampton University, Hampton, VA 23668}
\newcommand*{\NOWTulane }{ Tulane University, New Orleans, Lousiana  70118}
\newcommand*{\NOWKYUNGPOOK }{ Kyungpook National University, Daegu 702-701, South Korea}
\newcommand*{\NOWCUA }{ Catholic University of America, Washington, D.C. 20064}
\newcommand*{\NOWGEORGETOWN }{ Georgetown University, Washington, DC 20057}
\newcommand*{\NOWJMU }{ James Madison University, Harrisonburg, Virginia 22807}
\newcommand*{\NOWURICH }{ University of Richmond, Richmond, Virginia 23173}
\newcommand*{\NOWCALTECH }{ California Institute of Technology, Pasadena, California 91125}
\newcommand*{\NOWMOSCOW }{ Moscow State University, 119899 Moscow, Russia}
\newcommand*{\NOWVIRGINIA }{ University of Virginia, Charlottesville, Virginia 22901}
\newcommand*{\NOWYEREVAN }{ Yerevan Physics Institute, 375036 Yerevan, Armenia}
\newcommand*{\NOWRICE }{ Rice University, Houston, Texas 77005-1892}
\newcommand*{\NOWINFNGE }{ INFN, Sezione di Genova, 16146 Genova, Italy}
\newcommand*{\NOWBATES }{ MIT-Bates Linear Accelerator Center, Middleton, MA 01949}
\newcommand*{\NOWODU }{ Old Dominion University, Norfolk, Virginia 23529}
\newcommand*{\NOWVSU }{ Virginia State University, Petersburg,Virginia 23806}
\newcommand*{\NOWUNIONC }{ Department of Physics, Schenectady, NY 12308}
\newcommand*{\NOWORST }{ Oregon State University, Corvallis, Oregon 97331-6507}
\newcommand*{\NOWCNU }{ Christopher Newport University, Newport News, Virginia 23606}
\newcommand*{\NOWGWU }{ The George Washington University, Washington, DC 20052}
%%%%%%%%%%%%%%%%%%%% authors %%%%%%%%%   
\author{J.W.C.~McNabb}\affiliation{\CMU}\affiliation{\PENN}
\author{R.A.~Schumacher}\affiliation{\CMU}
\author{L.~Todor} \affiliation{\CMU}
\author{G.~Adams} \affiliation{\RPI}
\author{E.~Anciant}\affiliation{\SACLAY}
\author{M.~Anghinolfi}\affiliation{\INFNGE}
\author{B.~Asavapibhop}\affiliation{\UMASS}
\author{G.~Audit}\affiliation{\SACLAY}
\author{T.~Auger}\affiliation{\SACLAY}
\author{H.~Avakian}\affiliation{\JLAB}\affiliation{\INFNFR}
% ...
\collaboration{The CLAS Collaboration} \noaffiliation
%  
%The Southeastern Universities Research Association (SURA) operatesthe 
%Thomas Jefferson National Accelerator Facility for the United States 
%Department of Energy under contract DE-AC05-84ER40150.

\title{Hyperon Photoproduction in the Nucleon Resonance Region} 

\date{May 25, 2003}% It is always \today, but any date may be explicitly specified

\begin{abstract} 

Cross-sections and recoil polarizations for the reactions $\gamma + p
\rightarrow K^+ + \Lambda$ and $\gamma + p \rightarrow K^+ + \Sigma^0$
have been measured with high statistics and with good angular coverage
for center-of-mass energies between 1.6 and 2.3 GeV.  In the
$K^+\Lambda$ channel we confirm a structure near $W=1.9$ GeV at
backward kaon angles, but our data shows a more complex $s$- and $u$-
channel resonance structure than previously seen.  This structure is
present at forward and backward angles but not central angles, and its
position and width change with angle, indicating that more than one
resonance is playing a role.  Rising back-angle cross sections at
higher energies and large positive polarization at backward angles are
consistent with sizable $s$- or $u$-channel contributions.  None of
the model calculations we present can consistently explain these
aspects of the data.

\end{abstract}

 % end of PACS codes
\maketitle
%\pacs{
%      {13.30.-a}{ Decays of baryons},
%      {13.30.Eg}{ Hadronic decays},
%      {13.40.-f}{ Electromagnetic processes and properties},
%      {13.60.-r}{ Photon and charged lepton interactions with hadrons},
%      {13.60.Le}{ Meson production},
%      {14.20.Gk}{ Baryon resonances with S=0},
%      {25.20.Lj}{ Photoproduction reactions}
%     } % end of PACS codes
\maketitle

Characterizing the non-strange baryon resonances is of fundamental
interest in non-perturbative QCD. The masses, quantum numbers, and
decay branches of the higher-mass baryon resonances have remained
difficult to establish, both experimentally and theoretically.
Experimentally, most information comes from the use of pion beams
interacting with nucleon targets, combined with detection of one or
more pions or the nucleon.  For increasing masses, both the energy
overlap of resonances and meson production ({\it e.g.} the $\rho$)
make it more difficult to separate the resonance contributions.  The
long list of poorly-established higher-mass resonances~\cite{pdg}
illustrates this problem above the strangeness threshold near $W$=1600
MeV.  Theoretically, there is an apparent over-supply of baryons
predicted in quark models, the so-called ``missing baryons''
problem~\cite{cap}. Various ways have been suggested whereby
dynamical effects such as di-quarks could reduce the number of states to
something closer to what has been already observed~\cite{klempt}.

Photoproduction of non-strange resonances detected via decay into
strange particles offers two benefits in this field.  First, two-body
$KY$ final states are easier to analyze than the three-body $\pi\pi N$
final states that dominate decays at higher masses.  So, while the
cross sections for strangeness production tend to be small (on the
order of 1 or 2 $\mu$b in electromagnetic production), the energy and
angular distributions are simpler.  Also, the recoil polarization
observables are readily accessible via hyperon decays. Second,
couplings of nucleon resonances to $KY$ final states are expected to
differ from coupling to $\pi N$ or $\pi \pi N$ final
states~\cite{cap}.  Therefore, looking in the strangeness sector casts
a different light on the resonance excitation spectrum, and thus may
emphasize resonances not revealed in $\pi N$ scattering.  Some
``missing resonances'' may only be ``hidden'' by the character of the
channels studied previously.  To date, however, the PDG
compilation~\cite{pdg} gives poorly-known $K\Lambda$ couplings for
only five well-established resonances, and no $K\Sigma$ couplings for
any resonances. The most widely-available model calculation of the
$K\Lambda$ photoproduction, the Kaon-MAID code~\cite{maid}, includes a
mere three well-established resonances: the $S_{11}(1650)$, the
$P_{11}(1710)$, and the $P_{13}(1720)$.  Thus it is timely and
interesting to have additional good-quality photoproduction data of
these channels to see what additional resonance formation and decay
information can be obtained.  Here we report the global features of
our results~\cite{john} which are new, and compare them to published
reaction models.

Differential cross section and hyperon recoil polarization data were
obtained with the CLAS~\cite{clas0} system in Hall B at the Thomas
Jefferson National Accelerator Facility.  A beam of tagged photons
from a bremsstrahlung beam spanned energies from threshold at
$E_\gamma=0.911$ GeV ($W$ = 1.609 GeV) up to 2.325 GeV ($W$ = 2.290
GeV).  The event trigger required an electron signal from the photon
tagger, and at least one charged-track coincidence between the
time-of-flight `start' counters near the 18-cm liquid-hydrogen target
and the time-of-flight `stop' counters surrounding the drift chambers.
Kaons were identified using momentum and time-of-flight measurements
to compute their mass, and were the only particles detected in CLAS to
obtain the cross sections.  The $\Lambda$ and $\Sigma^0$ yields were
separated from the background due to mis-identified pions using
lineshape fits to missing-mass spectra in each of over 900 kinematic
bins of photon energy and kaon angle.  The results are binned in 25
MeV steps in $E_\gamma$ and in 18 bins in the center-of-mass angle of
the kaons, $-0.9 <\cos(\theta_K^{CM}) < +0.9$.  Consistency among
several variations of kaon selection cuts and background shapes was
demanded in extracting the hyperon yields, with $\chi^2_{\nu}$ always
less than 1.75 and signal-to-background ratios of greater than 2.5.  A
hyperon missing-mass resolution of $\sigma=6.1$ MeV was obtained when
averaged over all detection angles and photon energies.  The estimated
method-dependent yield uncertainties are included bin by bin in our
results, and average $6\%$.  A total of 427,000 $K^+ \Lambda$ events
and 354,000 $K^+ \Sigma^0$ events were accumulated.

The acceptance and efficiency for CLAS were modeled twice, using two
independent Monte Carlo models.  One was a full GEANT-based simulation
involving hit digitizations, while the other was a faster parametric
simulation that modeled detector effects starting at the level of
reconstructed tracks. The results were in good agreement overall, and
analysis of the remaining variations led to an estimated global
systematic uncertainty of $7\%$.  This was the dominant systematic
uncertainty in the experiment.

The photon flux was determined by integrating the tagger rate.  The
rate was sampled by counting hits from accidental photons in the
tagger TDC's.  Photon losses due to beam collimation were determined
using a separate total-absorption counter downstream of CLAS.  As a
check on our results, the $p(\gamma,\pi^+)n$ cross section was
measured using the same analysis chain as the $p(\gamma,K^+)Y$ data.
The pion cross section was found to be in good agreement with the
SAID~\cite{said} parameterization of the world's data between 0.6 and
1.6 GeV, albeit with an overall scale factor of 0.92.  In energy and
angle the variation in the ratio of CLAS to SAID pion cross sections
was small, approximately $\pm3\%$, compared to our overall kaon cross
section uncertainties.  This shows that the yield extractions,
acceptance calculations, and photon flux determinations were all
consistent, but the overall normalization was made to the world's pion
photoproduction data.  The final global systematic uncertainties on
our cross sections are $8.2\%$ for the $\Lambda$ data and $7.7\%$ for
the $\Sigma^0$ data.

\begin{figure}
\resizebox{0.50\textwidth}{!}{\includegraphics{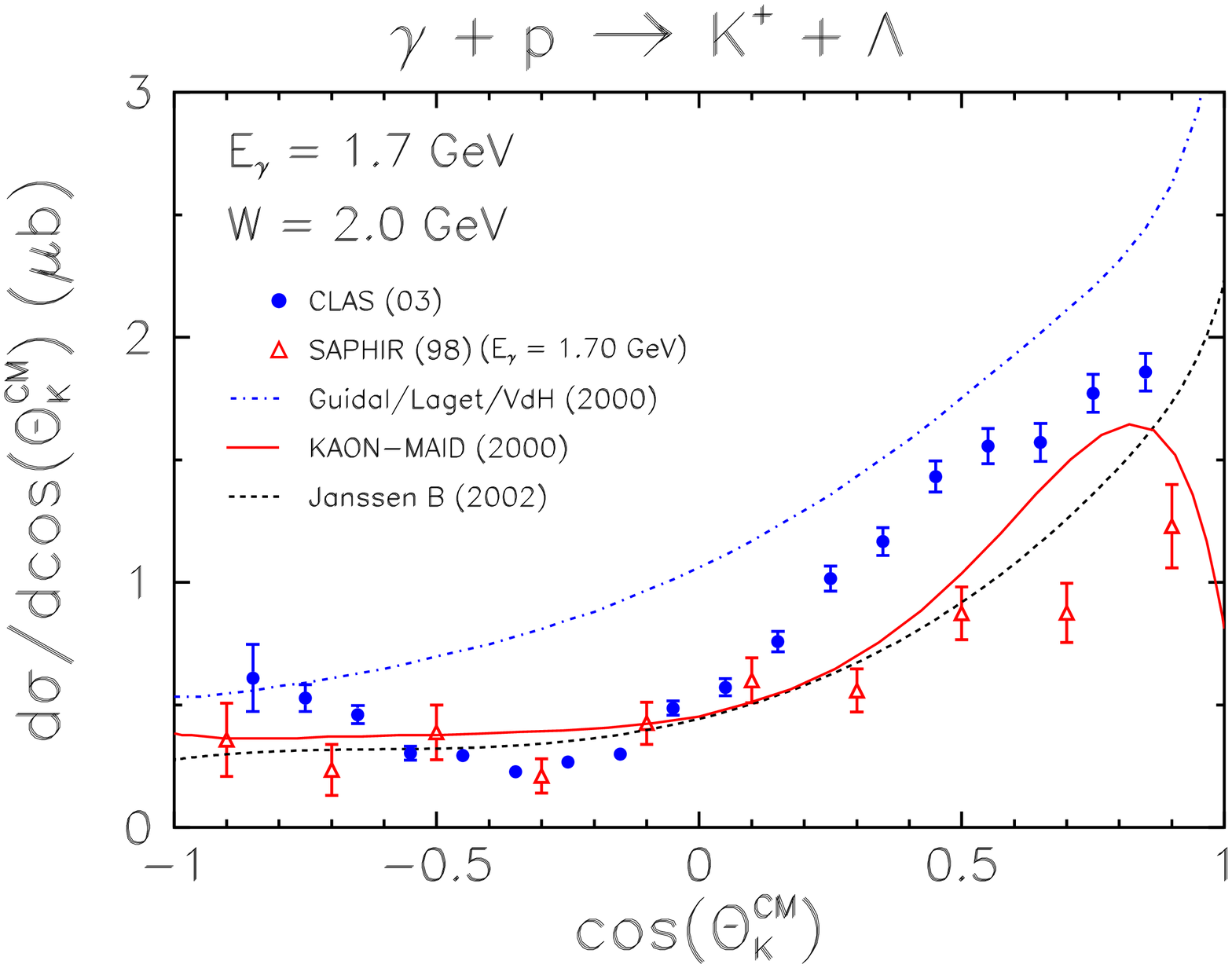}}
\resizebox{0.50\textwidth}{!}{\includegraphics{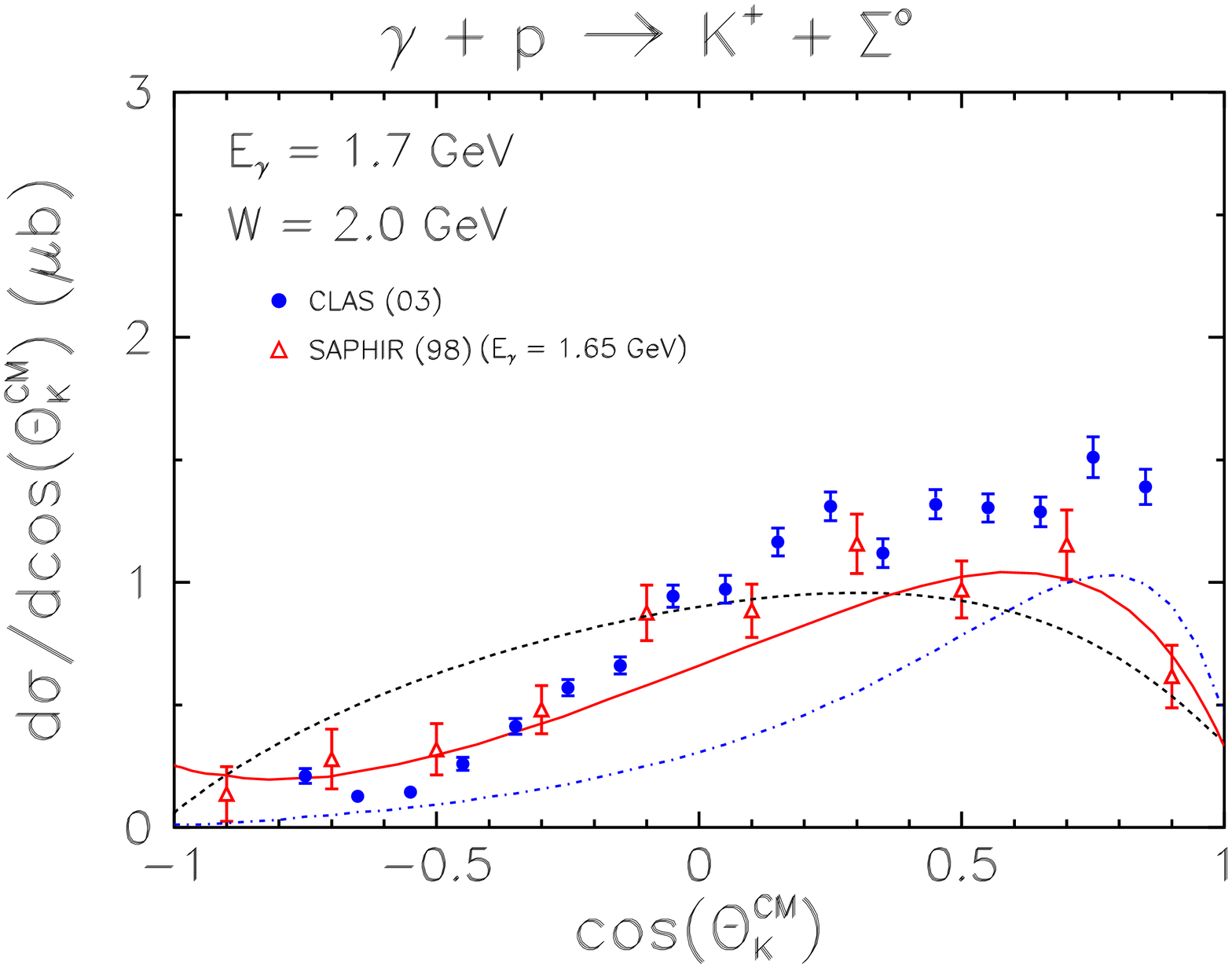}}
%\vspace{5cm}       % Give the correct figure height in cm
\caption{
Angular distributions for $\Lambda$ (top) and $\Sigma^0$ (bottom) 
hyperon photoproduction
measured at CLAS (solid circles) at W=2.01 GeV. Data from
{\small SAPHIR}~\cite{bon} (open triangles) are also shown.  The
curves are for effective Lagrangian calculations 
computed by Kaon-MAID~\cite{maid} (solid), the
alternative model of Janssen {\it et al.}~\cite{jan} (dashed), and a
Regge-model calculation of Guidal {\it et al.}~\cite{lag1,lag2}
(dot-dashed).}
\label{fig:a}       % Give a unique label
\end{figure}

Figure \ref{fig:a} (top) shows the differential cross section for
$\Lambda$ hyperon photoproduction at $W = 2.0$ GeV.  It is forward
peaked, as has been seen in previous experiments~\cite{bon}.  However, we
also see a backward rise in the cross section for this and similar
high values of $W$.  This can be due either to $u$-channel components
of the reaction mechanism or to the interference of $s$-channel
resonances.  The agreement between CLAS and previous data from {\small
SAPHIR} at Bonn~\cite{bon} varies: generally the measurements agree
within the estimated uncertainties at back angles and near threshold
energies, but CLAS measures consistently larger $K^+\Lambda$ cross
sections at forward kaon angles.

The $\Lambda$ and $\Sigma^0$ hyperons have isospin 0 and 1,
respectively, and so intermediate states leading to the production of
$\Lambda$'s can only have isospin 1/2 ($N^*$ only), whereas for the
$\Sigma^0$'s intermediate states with both isospin 1/2 and 3/2 ($N^*$
or $\Delta$) can contribute.  Figure~\ref{fig:a} (bottom) shows the
data for $\Sigma^0$ production at the same $W$ as above, showing
different trends induced by differing resonance structure.  In general
the $\Sigma^0$ cross sections are less forward peaked, with less
indication of a back-angle rise.

%%%%%%%%%%%%%%%%%%%%%%%%%%%%%%%%%%%%%%%%%%%%%%%%%%%%%%%%%%%%%%
%%%%%%%%%%%%%%%%%%%%%%%%%%%%%%%%%%%%%%%%%%%%%%%%%%%%%%%%%%%%%%

The Regge-model calculation~\cite{lag1,lag2} shown in
Figure~\ref{fig:a} uses only $K$ and $K^*$ exchanges, with no
$s$-channel resonances.  The prediction was made using a model that
fit high-energy kaon electroproduction data well, and could be
expected to reproduce the average behavior of the cross section in the
resonance region.  However, extrapolated down to the resonance region,
the model overpredicts the size of the $\Lambda$ cross section and
underpredicts that of the $\Sigma^0$.  Since it is a $t$-channel
reaction model, it cannot produce a rise at back angles as
seen for the $\Lambda$, and illustrates the need for $s$- and
$u$-channel contributions to understand that feature.  Two
hadrodynamic models based on similar effective Lagrangian
approaches~\cite{maid,mart}~\cite{jan} are also shown.  Both emphasize the
addition of a small set of $s$-channel resonances to the
non-resonant Born terms, and differ in their treatment of hadronic
form factors and gauge invariance restoration.  Both were fit to the
previous data from {\small SAPHIR}\cite{bon}, and therefore do not
agree well with our results.

\begin{figure}
\resizebox{0.50\textwidth}{!}{\includegraphics{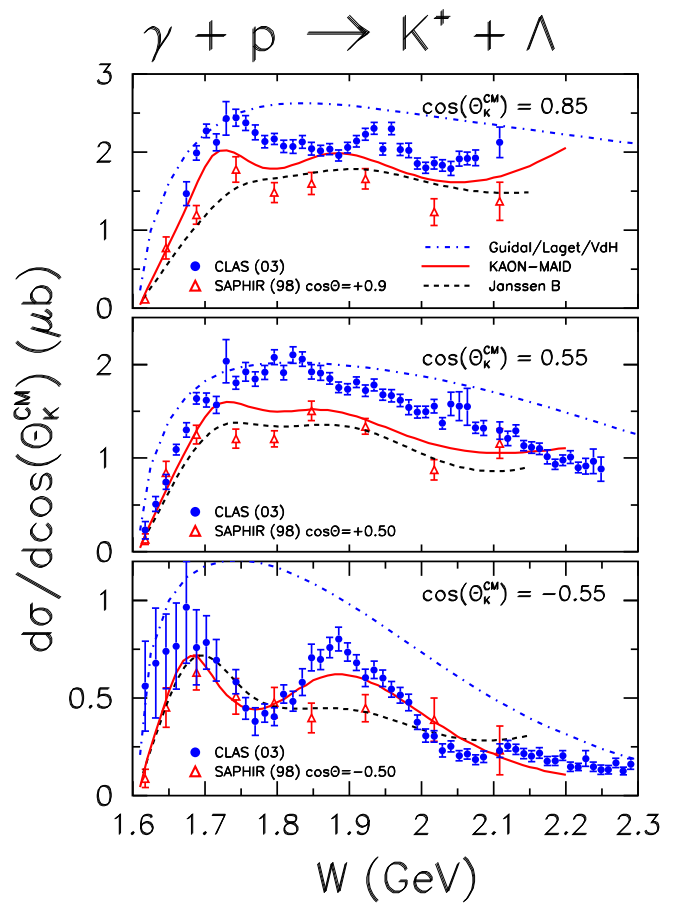}}
\caption{Energy dependence of the $\Lambda$ cross section at the most
forward angle measured (top), and at intermediate forward and backward
angles (middle, bottom).}
\label{fig:b}       % Give a unique label
\end{figure}

Resonance structure in the $s$-channel should appear most clearly in
the $W$ dependence of the cross sections.  In Figure~\ref{fig:b} (top)
we show the $K^+ \Lambda$ cross section at our most forward kaon
angle, showing a sharp rise from threshold up to 1.72 GeV, a slow
decline, and then a structure at 1.95 GeV with a full-width of about
100 MeV.  The peak in the threshold region is understood in model
calculations as due to the known $S_{11}(1650)$, $P_{11}(1710)$, and
$P_{13}(1720)$ resonances.  At a moderate forward angle, shown in
Figure~\ref{fig:b} (middle), the higher mass structure near 1.95 GeV
is not visible.  At a moderate backward angle, shown in
Figure~\ref{fig:b} (bottom), we again see clear structure, but it is
broader, centered near 1.90 GeV, and is about 200 MeV wide.  These
structures are prominent at forward and backward angles; for most
intermediate angles the energy dependence near 1.9 GeV falls smoothly.
In contrast, our corresponding measured $\pi^+n$ cross sections are
featureless throughout this angle and energy range~\cite{john}.

Indications of the structure near 1.9 GeV were first seen in data from
{\small SAPHIR}~\cite{bon}, which was interpreted by some~\cite{mart}
as evidence for a ``missing'' resonance at this mass.  Based on
theoretical guidance from one particular quark model~\cite{cap}, an
assignment of $D_{13}(1895)$ seemed consistent with the angular
distributions.  However, other groups~\cite{sag,jan} showed that the
same data could be accommodated using $u$-channel hyperon exchanges,
an extra $P$-wave resonance, or alternative hadronic form factors.
CLAS data, which shows a structure that varies in width and position
with kaon angle, suggests an interference phenomenon between several
resonant states in this mass range, rather than a single
well-separated resonance.  This should be expected, since many
$s$-resonances occupy this mass range. The best modeling of the
backward-angle structure near 1.9 GeV is given in Ref.~\cite{mart} by
incorporating a $D_{13}(1895)$. We see, however, that the resulting
fixed position and width in this model is not consistent with the
variation with angle seen in the data.

%%%%%%%%%%%%%%%%%%%%%%%%%%%%%%%%%%%%%%%%%%%%%%%%%%%%%%%%%%%%%%
%%%%%%%%%%%%%%%%%%%%%%%%%%%%%%%%%%%%%%%%%%%%%%%%%%%%%%%%%%%%%%

The hyperon recoil polarization provides another test of reaction
models.  This observable is related to interferences of the imaginary
parts of the resonant amplitudes with the real part of other
amplitudes, including the non-resonant Born terms.  Unpolarized
photons on an unpolarized target can only produce hyperons that are
polarized along the axis $(\hat{\gamma}\times\widehat{K^{+}})$
normal to the production plane.  The parity-violating weak decay
asymmetry in hyperon decays enables us to determine this polarization
by measuring the angular distribution of the decay protons.  The large
acceptance of CLAS made it straightforward to detect protons from the
decay of hyperons in coincidence with the $K^+$ mesons.

\begin{figure}
\resizebox{0.50\textwidth}{!}{\includegraphics{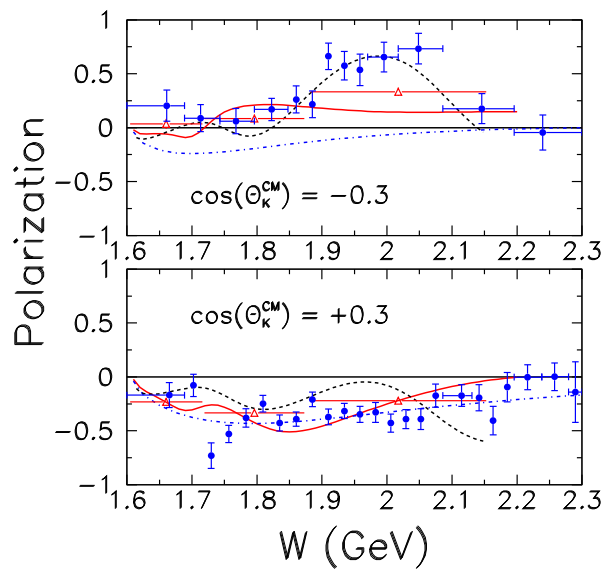}}
\vspace{-1.0cm}       % Give the correct figure height in cm
\caption{
Recoil polarization of $\Lambda$ hyperons as a function of $W$ for the
center-of-mass kaon angle of $\cos(\theta_{K}^{CM})= -0.3$ (top) and
$\cos(\theta_{K})= +0.3$ (bottom).  Vertical bars on CLAS data (solid
points) combine statistical and systematic errors and horizontal
error bars span regions of weighted averaging.  The curves and other
data are the same as in Figures~\ref{fig:a} and ~\ref{fig:b}.}
\label{fig:c}       % Give a unique label 
\end{figure}

Figure~\ref{fig:c} shows the $\Lambda$ recoil polarization as a
function of $W$ for representative kaon angles in the backward and the
forward directions. The data has been binned such that the statistical
uncertainty on each data point is less than $\pm0.15$.  The error bars
combine statistical and estimated systematic uncertainties arising
from the yield extraction.  Our results are generally consistent with
a few older data points from {\small SAPHIR}~\cite{bon}, but our
energy binning is finer and reveals more structure.  The data shows
negative polarization of the $\Lambda$ hyperons when the kaons go
forward in the center-of-mass frame and a comparably strong positive
polarization when the kaons go backward.

Of the three models tested here, only the Janssen {\it et
al.}~\cite{jan} model (dashed line) predicts the large back-angle
polarization seen in the data near 2.0 GeV.  This prediction is
strongly influenced by $u$-channel $Y^*$ contributions in that model
that are added to a $D_{13}(1895)$ $s$-channel component.  At
the forward angle, however, this model does not perform better than
the other hadrodynamic calculation or the Regge-based model.  The
positive back-angle polarization arises in the Kaon-MAID~\cite{maid}
calculation from the presence of a $D_{13}(1895)$, but the strength is
too small.  The Regge-based model~\cite{lag1,lag2} can provide only
very weak back-angle polarization, since by construction it has only a
$t$-channel (forward-angle) production mechanism, leading here to
the wrong sign.  Near threshold energy the hadrodynamic models show
the most structure due to interference of the known resonances cited
earlier; but here our data has limited precision and cannot
distinguish among these models.

%%%%%%%%%%%%%%%%%%%%%%%%%%%%%%%%%%%%%%%%%%%%%%%%%%%%%%%%%%%%%%
%%%%%%%%%%%%%%%%%%%%%%%%%%%%%%%%%%%%%%%%%%%%%%%%%%%%%%%%%%%%%%

In summary, we present results from an experimental investigation of
hyperon photoproduction from the proton in the energy range where
nucleon resonance physics should dominate.  Our $K^+\Lambda$ cross
section data reveals an interesting $W$-dependence: double-peaked at
forward and backward angles, but not at central angles.  The structure
near 1.9 GeV shifts in position and shape from forward to backward
angles.  This finding cannot be explained by a $t$-channel Regge-based
model or by the addition of a single new resonance in the $s$ or $u$
channel.  Our polarization data shows large values of polarization that
change from negative values at forward angles to positive for backward
kaon angles.  Since a $t$-channel Regge model is unable to explain the
backward, positive polarization, it appears that additional $s$- or
$u$-channel resonances are needed to explain the data.  Our results
show that hyperon photoproduction can reveal resonance structure
previously ``hidden'' from view.  Comprehensive partial wave analysis
and amplitude modeling for this data can therefore be hoped to firmly
establish the mass and possibly the quantum numbers of these states.

We thank staff of the Accelerator and the Physics Divisions at JLab.
Major support came from the U.S. DOE and NSF, INFN Italy, the French
atomic energy agency, and the Korean S\&E Foundation.  The
Southeastern Universities Research Association (SURA) operates the
Jefferson Lab for the United States DOE under contract
DE-AC05-84ER40150.

\end{document}